\documentclass[english]{revtex4}
\usepackage[T1]{fontenc}
\usepackage[latin9]{inputenc}
\usepackage{amsmath}
\usepackage{graphicx}
\usepackage{amssymb}

\makeatletter
\def\DR{\rm I\kern-1.45pt\rm R}\def\DC{\kern2pt {\hbox{\sqi I}}\kern-4.2pt\rm C}\newcommand{\ba}{\begin{array}}\newcommand{\ea}{\end{array}}\newcommand{\be}{\begin{equation}}\newcommand{\ee}{\end{equation}}\newcommand{\bea}{\begin{eqnarray}}\newcommand{\eea}{\end{eqnarray}}\newcommand{\bi}{\begin{itemize}}\newcommand{\ei}{\end{itemize}}

\usepackage{amscd}

\makeatother
\usepackage{babel}

\begin{document}
\title{Exactly solvable mixed-spin Ising-Heisenberg diamond chain with the biquadratic interactions and single-ion anisotropy}
\author{Onofre Rojas$^{1}$\footnote{email: ors@dex.ufla.br}, S. M. de Souza$^{1}$, Vadim Ohanyan$^{2,3}$\footnote{email: ohanyan@yerphi.am} and Martiros Khurshudyan$^{2}$}
\affiliation{$\;^{1}$Departamento de Ciencias Exatas, Universidade Federal de Lavras, CP 3037, 37200000, Lavras, MG, Brazil.}
\affiliation{ $\;^{2}$Yerevan State University, A.Manoogian, 1, Yerevan, 0025 Armenia.}
\affiliation{ $\;^{3}$Yerevan Physics Institute, Alikhanian Br.2, Yerevan, 0036, Armenia.}

\begin{abstract}
An exactly solvable variant of mixed spin-(1/2,1) Ising-Heisenberg diamond chain is considered. Vertical spin-1 dimers are taken as
quantum ones with Heisenberg bilinear and biquadratic interactions
and with single-ion anisotropy, while all interactions between spin-1
 and spin-1/2  residing on the intermediate sites are taken
in the Ising form. The detailed analysis of the $T=0$ ground state phase
diagram is presented.
The phase diagrams have shown to be rather rich, demonstrating large
variety of ground states: saturated one, three ferrimagnetic with
magnetization equal to 3/5 and another four ferrimagnetic ground states
with magnetization equal to 1/5. There are also two frustrated macroscopically
degenerated ground states which could exist at zero magnetic filed.
 Solving the model exactly within classical transfer-matrix
formalism we obtain an exact expressions for all thermodynamic function
of the system. The thermodynamic properties of the model have been described exactly
by exact calculation of partition function within the direct classical
transfer-matrix formalism, the entries of transfer matrix, in their
turn, contain the information about quantum states of vertical spin-1
XXZ dimer (eigenvalues of local hamiltonian for vertical link).
\end{abstract}
\maketitle
\section{Introduction}

\begin{figure}
\includegraphics[scale=0.8]{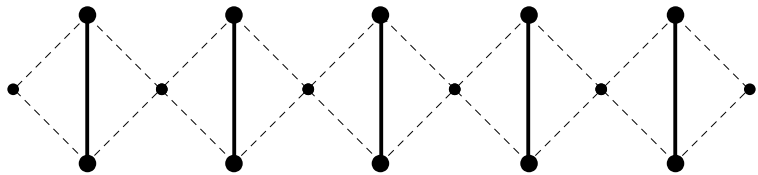}
\caption{\label{Fig-lat} The diamond chain with Ising and Heisenberg
bonds. Solid bold lines denote XXZ quantum bonds, dotted lines
correspond to Ising interactions. Large(small) circles denote
$\mathbf{S}$ $(\sigma)$ spins.}
\end{figure}

Lattice models of quantum magnetism continue to be in the focus of
attention of theoretical condensed matter physicists. Beside of great
practical importance connected with description of magnetic and thermodynamic
properties of real magnetic materials, this research area is also
attractive from the general statistical mechanics and strongly correlated
system theory points of view, especially when one deals with exactly
solvable strongly interacting many-body system. The diamond chain
is a one-dimensional lattice spin system in which the vertical spin
dimers alternate with single spins(See figure \ref{Fig-lat}). This
model with $S=1/2$ is believed to describe the magnetic lattice of
mineral azurite, Cu$_{3}$(CO$_{3}$)$_{2}$(OH)$_{2}$, which is
famous for its deep blue pigmentation\cite{az1,az2,az3,az4}. Theoretical
research of various aspects of diamond chain physics received much
attention during last years\cite{dc1}-\cite{dcm3}. Diamond chain
and especially diamond chain with mixed $(S,S/2)$ spin are shown
to have very rich ground state phase diagrams with Haldane and several
spin-cluster states which are tensor product of exact local eigenstates
of cluster spins \cite{dc1,dc2}. Many other issues of diamond chain
physics have been investigated theoretically during last years including
Dzyaloshinsky-Moriya term influence on magnetization processes\cite{dc3},
multiple-spin-exchange effects\cite{dc4}, magnetization plateaus\cite{fu,dc5},
magnetocaloric effect\cite{dc6}, e.t.c. Very recently another one
interesting feature of diamond chain, possibility of localized magnon
excitations, has been also investigated\cite{hon}.

Especially important issue is the effect of frustration, which is
rather strong in antiferromagnetic diamond chain because of triangular
arrangement of the sites. Variants of frustrated recurrent lattices
with diamond plaquette have been also studied in Refs. \cite{kob09}
and \cite{ara03}. However, diamond chain is not integrable in general.
Thus, the exact analysis of the dynamic and especially thermodynamic
properties of diamond chain is a very complicated issue. Nevertheless,
one can consider various exactly solvable variants of spin systems
possessing diamond chain topology of interaction bonds with simplified
structure of interactions\cite{dci}-\cite{dcm3}. The diamond chain
with only Ising type of interaction have been exactly solved within
classical transfer-matrix technique in Ref. \cite{dci} revealing
reach $T=0$ ground state phase diagram. Yet another exactly solvable
diamond chains have been considered in Refs. (\cite{dcm1}-\cite{dcXX}),
where vertical spin dimers have been taken as quantum ones with $XXZ$
interaction while interaction between spins localized on vertical
dimer sites and spins from the single sites alternating with them
is of Ising type. Mapping the system into a single Ising chain within
iteration-decoration transformation\cite{Fisher,syozi,physA388} the
authors gave complete description of the ground state properties,
$T=0$ ground state phase diagram as well as thermodynamic functions
for all values of vertical dimer spins magnitude $S$. In a very recent
paper the diamond chain with $XX$-interaction has been considered
in the Jordan--Wigner formalism\cite{dcXX}. %

Considering the mixed spin chains (or another one-dimensional spin
systems with more complicated geometry) with Ising and Heisenberg
bonds (or even just Ising counterparts of the known quantum spin models)
one can achieve two-fold goal: to construct novel exactly solvable
lattice spin model which allow one to obtain analytic expression for
all thermodynamic functions of the model, and to get an approximate
models which can be useful for understanding the properties of underlying
purely quantum models \cite{dcm1,dcm2,dcm3,roj08,hov08,ant09,oha10,oha09,bel09,str05}.
Exact thermodynamic solutions of such models even can shed a light
to the properties of real magnetic materials. For instance, for alternating
spin chain even simplest models with only Ising interaction can reflect
the underlying magnetic behavior of the corresponding Heisenberg counterpart
at least in qualitative way \cite{oha03,lit08}, moreover, some exactly
solvable models with Ising and Heisenberg bonds can also provide satisfactory
quantitative picture\cite{str05}. Very recently, the synthesis of
novel class of trimetalic {3d-4d-4f} coordination polymers has been
reported. One of them, a 1d coordination polymer compound containing 3d (Cu$^{2+}$), 4d (Mo$^{5+}$) and 4f (Dy$^{3+}$)
ions is shown to exhibit the properties of 1d magnet with Ising and Heisenberg bonds\cite{chem}. The appearance of
Ising interactions between magnetic ion in this compound is connected with the extremely anisotropic properties
of Dy$^{3+}$ ground states ($g_{\|}=19.6$, $g_{\bot}\approx0$). Thus, the interactions of Dy$^{3+}$ with
the surrounding Cu$^{2+}$ and  Mo$^{5+}$ ions are, to the great extent, of Ising type, involving spin projection
along the dysprosium anisotropy axis. While, the interaction bonds Cu$^{2+}$---Mo$^{5+}$ correspond to Heisenberg
interaction\cite{chem}.Though, the aforementioned 1d coordination polymer system is not exactly the Ising--Heisenberg
diamond chain considered in the present paper, this discovery of novel classes of magnetic materials  makes the investigation
of exact solutions of spin chains with Ising and Heisenberg
bonds important form practical point of view as well.

In this paper we consider mixed spin-(1,1/2) diamond chain with Ising
and Heisenberg bonds which extends the system considered in Ref. \cite{dcm3}
by including biquadratic term for $S=1$ $XXZ$-dimers and single-ion
anisotropy. Biquadratic terms are usually originated from the spin-lattice
coupling in the adiabatic phonons approximation\cite{kit} but also
can be considered as the effect of quadrupole interaction between
the spins. We do not make any assumption about the origin of biquadratic
terms considering them as a part of general Hamiltonian. The model
allow one to calculate the partition function and, thus, all thermodynamic
quantities exactly within classical transfer-matrix formalism\cite{bax}.
We present the analysis of $T=0$ ground-state phase diagrams and
plot the curves of magnetization processes for finite temperatures,
demonstrating magnetization plateaus at $M=1/5$ and $M=3/5$ in the
units of saturation magnetization.

The paper is organized as follows. In the second Section we
formulate the model and present eigenvalues of block Hamiltonian. In
the third section we describe possible ground states of the system
and present the ground-states phase diagram. In section 4 we present
its exact solution and discuss the magnetization and thermodynamics
properties of the model. Finally in sec. 5 a short summary is
followed.

\section{The model and its exact solution}

Let us consider the system of vertical $S=1$ spin dimers with Heisenberg
$XXZ$ bilinear and biquadratic interactions and uniaxial single-ion
anisotropy. These dimers are assembled to the chain by alternating
with Ising spins $\sigma$, so that each spin $S$ in certain dimer
interacts to its both left and right Ising spins via Ising type interaction
(See figure \ref{Fig-lat}). So, we have the so--called diamond--chain
with $S=1$ Heisenberg dimers and $\sigma=1/2$ Ising spins between
them. The corresponding Hamiltonian is the sum over the block Hamiltonians:

\begin{eqnarray}
 &  & \mathcal{H}=\sum_{i=1}^{N}(\mathcal{H}_{i}-\frac{h_{2}}{2}(\sigma_{i}+\sigma_{i+1})),\\
 &  & \mathcal{H}_{i}=J(\mathbf{S}_{i1}\cdot\mathbf{S}_{i2})_{\Delta}+K(\mathbf{S}_{i1}\cdot\mathbf{S}_{i2})_{\Delta}^{2}+D\left((S_{i1}^{z})^{2}+(S_{i2}^{z})^{2}\right)+J_{0}(\sigma_{i}+\sigma_{i+1})(S_{i1}^{z}+S_{i2}^{z})-h_{1}(S_{i1}^{z}+S_{i2}^{z}),\nonumber \end{eqnarray}
 where $N$ is the number of the unit cells (blocks with two spin-1
and one spin-1/2), while $i$ correspond the particles at $i$-cell.
$J$ being the coupling constant of bilinear $XXZ$-interaction between
Heisenberg spins, which we assume to be of the following form

\begin{equation}
(\mathbf{S}_{i1}\cdot\mathbf{S}_{i2})_{\Delta}=\Delta(S_{i1}^{x}S_{i2}^{x}+S_{i1}^{y}S_{i2}^{y})+S_{i1}^{z}S_{i2}^{z},\end{equation}
 whereas $K$ denotes the biquadratic XXZ-interaction term, $D$ means
the single ion-anisotropy and $J_{0}$ being the purely Ising interaction
term. $h_{2}$ and $h_{1}$ are the external magnetic field acting
on $\sigma_{i}$ and $S_{i}$ respectively.

In order to solve this model, at first, we need to diagonalize the
block Hamiltonian for arbitrary $i$-th block. Nine eigenvalues of
$\mathcal{H}_{i}$, $\lambda_{n}(\sigma_{i},\sigma_{i+1})$, $n=1...9$
can be found analytically, which write down as

\begin{eqnarray}
\lambda_{1,2} & = & J+K+2D\pm 2[-h_{1}+J_{0}(\sigma_{i}+\sigma_{i+1})],\nonumber \\
\lambda_{3,4} & = & \Delta(J+\Delta K)+D\pm [-h_{1}+J_{0}(\sigma_{i}+\sigma_{i+1})],\nonumber \\
\lambda_{5,6} & = & -\Delta(J-\Delta K)+D\pm [-h_{1}+J_{0}(\sigma_{i}+\sigma_{i+1})],\nonumber \\
\lambda_{7} & = & -J+K+2D,\nonumber \\ \label{eig}
\lambda_{8,9} & = & \frac{-J+(1+4\Delta^{2})K+2D}{2}\pm\frac{1}{2}R,
\end{eqnarray}
where for simplicity $R$ denotes the following expression,,
 \begin{equation}
R=\sqrt{8\Delta^{2}(J-K)^{2}+(J-K-2D)^{2}}.\label{eq.R}
\end{equation}

The eigenvectors of block Hamiltonian $\mathcal{H}_{i}$ up to the
inversion of all spins are

\begin{align}
|v_{2}\rangle= & |1,1\rangle, &  & \Rightarrow\qquad\lambda_{1},\lambda_{2}\nonumber \\
|v_{1,s}\rangle= & \frac{1}{\sqrt{2}}(|1,0\rangle+|0,1\rangle), &  & \Rightarrow\qquad\lambda_{3},\lambda_{4}\nonumber \\
|v_{1,a}\rangle= & \frac{1}{\sqrt{2}}(-|1,0\rangle+|0,1\rangle), &  & \Rightarrow\qquad\lambda_{5},\lambda_{6}\nonumber \\
|v_{0,a}\rangle= & \frac{1}{\sqrt{2}}(-|-1,1\rangle+|1,-1\rangle), &  & \Rightarrow\qquad\lambda_{7}\nonumber \\
|v_{0,\pm}\rangle= & \frac{1}{\sqrt{2+c_{\pm}^{2}}}(|-1,1\rangle+c_{\pm}|0,0,\rangle+|1,-1\rangle), &  & \Rightarrow\qquad\lambda_{8},\lambda_{9}\label{eq:egst}\end{align}
 where $1,0$ and $-1$ in first(second) place stand for the $S^{z}=1,0$
and $-1$ states for first(second) spin in vertical dimer and the
following notation is adopted:

\begin{equation}
c_{\pm}=\frac{1}{2\Delta}\left(1+\frac{2D\pm R}{K-J}\right).\label{eq:cpm}\end{equation}

The eigenvectors of the dimer defined by $|v_{2}\rangle$ corresponds
to the parallel ordered spins with magnetization per site $m_{s}=1$,
the eigenvectors $|v_{1,s}\rangle$ and $|v_{1,a}\rangle$ correspond
to symmetric and antisymmetric states vector respectively, whereas
$|v_{0,a}\rangle$ is an antisymmetric state vector with
magnetization $m_{s}=0$, finally $|v_{0,\pm}\rangle$ are the
symmetric eigenvectors with magnetization $m_{s}=0$.

The remaining eigenvectors of the eigenvalues $\lambda_{2}$, $\lambda_{4}$
and $\lambda_{6}$ can be obtaining using the spin inversion.

\subsection{Special case $K=J$}

At the special value $K=J$ the $S_{tot}^{z}=0$ sector of the block
Hamiltonian undergoes qualitative changes which can be obtained substituting
carefully $K=J$ value into the general solution presented above in
eq.\eqref{eq:egst}. This should be considered as a consequence of
the special symmetry of the Hamiltonian for these values of parameters,
more precisely, the fact that operator $(\mathbf{S}_{1}\cdot\mathbf{S}_{2})_{\Delta}+(\mathbf{S}_{1}\cdot\mathbf{S}_{2})_{\Delta}^{2}$
can be represented in terms of permutation operators $P_{12}$. In
this case the eigenstates $|v_{0,a}\rangle$ and $|v_{0,\pm}\rangle$
of the Hamiltonian reduce to the following ones

\begin{align}
|v_{0,a}\rangle= & \frac{1}{\sqrt{2}}(|1,-1\rangle-|-1,1\rangle), &  & \Rightarrow\qquad\lambda_{7}=2D,\nonumber \\
|v_{0,+}\rangle= & |0,0\rangle, &  & \Rightarrow\qquad\lambda_{8}=2J\Delta^{2},\nonumber \\
|v_{0,-}\rangle= & \frac{1}{\sqrt{2}}(|1,-1\rangle+|-1,1\rangle), &  & \Rightarrow\qquad\lambda_{9}=2(J\Delta^{2}+D).\end{align}

Note that a straightforward substitution in eq.\eqref{eq:cpm} could
yield to an undefined coefficients of the eigenstates.

\section{Ground states phase diagrams}

Let us describe possible $T=0$ ground states of the chain under
consideration and the corresponding energies per block. Generally
speaking, there are $9\times2=18$ possible ground states for each
block. However, if one restricts himself with the ground states
which are equivalent up to the inversion of all spins one will
arrive at the following spin configurations. The fully polarized
state ($M=1$, here $M$ is magnetization per spin) \begin{eqnarray}
|SP\rangle=\prod_{i=1}^{N} &
|v_{2}\rangle_{i}\otimes|\uparrow\rangle_{i},\quad\varepsilon_{SP}=J+K+2D+2J_{0}-\frac{5}{2}H,\label{eq:SP}\end{eqnarray}
 where $|\uparrow\rangle_{i}$($|\downarrow\rangle_{i}$) stands for
the up(down) state of the $\sigma$-spin in the $i$-th block. Hereafter,
we also put $h_{1}$=$h_{2}$=$H$. The next sector of ground states
contains three different ferrimagnetic spin configurations with the
value of magnetization equal to 3/5 ($M=3/5$), \begin{eqnarray}
 &  & |F1\rangle=\prod_{i=1}^{N}|v_{2}\rangle_{i}\otimes|\downarrow\rangle_{i},\quad\varepsilon_{F1}=J+K+2D-2J_{0}-\frac{3}{2}H,\nonumber \\
 &  & |F2\rangle=\prod_{i=1}^{N}|v_{1,s}\rangle_{i}\otimes|\uparrow\rangle_{i},\quad\varepsilon_{F2}=\Delta(J+\Delta K)+D+J_{0}-\frac{3}{2}H,\nonumber \\
 &  & |F3\rangle=\prod_{i=1}^{N}|v_{1,a}\rangle_{i}\otimes|\uparrow\rangle_{i},\quad\varepsilon_{F3}=-\Delta(J-\Delta K)+D+J_{0}-\frac{3}{2}H.\label{eq:F1-3}\end{eqnarray}
 There are also four another ferrimagnetic ground states with $M=1/5$:

\begin{eqnarray}
 &  & |F4\rangle=\prod_{i=1}^{N}|v_{1,s}\rangle_{i}\otimes|\downarrow\rangle_{i},\quad\varepsilon_{F4}=\Delta(J+\Delta K)+D-J_{0}-\frac{1}{2}H,\nonumber \\
 &  & |F5\rangle=\prod_{i=1}^{N}|v_{1,a}\rangle_{i}\otimes|\downarrow\rangle_{i},\quad\varepsilon_{F5}=-\Delta(J-\Delta K)+D-J_{0}-\frac{1}{2}H,\nonumber \\
 &  & |F6\rangle=\prod_{i=1}^{N}|v_{0,a}\rangle_{i}\otimes|\uparrow\rangle_{i},\quad\varepsilon_{F6}=-J+K+2D-\frac{1}{2}H,\nonumber \\
 &  & |F7\rangle=\prod_{i=1}^{N}|v_{0,-}\rangle_{i}\otimes|\uparrow\rangle_{i},\quad\varepsilon_{F7}=\frac{1}{2}\left(-J+(1+4\Delta^{2})K+2D-R\right)-\frac{1}{2}H.\label{eq:F4-7}\end{eqnarray}

When no external magnetic field is applied, there are also possibility
of frustrated ground state formation, in which the orientation of
$\sigma$ spins in each block are not defined. In this case all $\sigma$
spins become decoupled and behave like free spins. There are two frustrated
ground states \begin{eqnarray}
 &  & |FR1\rangle=\prod_{i=1}^{N}|v_{0,-}\rangle_{i}\otimes|\xi\rangle_{i},\quad\varepsilon_{FR2}=\frac{1}{2}\left(-J+(1+4\Delta^{2})K+2D-R\right),\nonumber \\
 &  & |FR2\rangle=\prod_{i=1}^{N}|v_{0,a}\rangle_{i}\otimes|\xi\rangle_{i},\quad\varepsilon_{FR1}=-J+K+2D.\label{eq:frus}\end{eqnarray}

Here $|\xi\rangle_{i}$ stands for arbitrary value of the $\sigma$
spin in the $i$-th block. Thus, for $H=0$ if $S=1$ dimer has
$S_{tot}^{z}=0$ then its neighboring $\sigma$-spins become decoupled
which results in macroscopic non-zero entropy $S/N=\log2$ for each
of the frustrated ground states of Eq. \ref{eq:frus}. Applying
magnetic field one removes the two-fold macroscopic degeneracy
driving $|FR1\rangle$ and $|FR2\rangle$ ground states into
$|F7\rangle$ and $|F6\rangle$ respectively. Nevertheless, in some
papers two last non-degenerated ground states of eqs.\eqref{eq:F4-7}
are mentioned as frustrated ones \cite{dcm1,dcm2,dcm3}.

\begin{figure}
\includegraphics[scale=0.7,angle=-90]{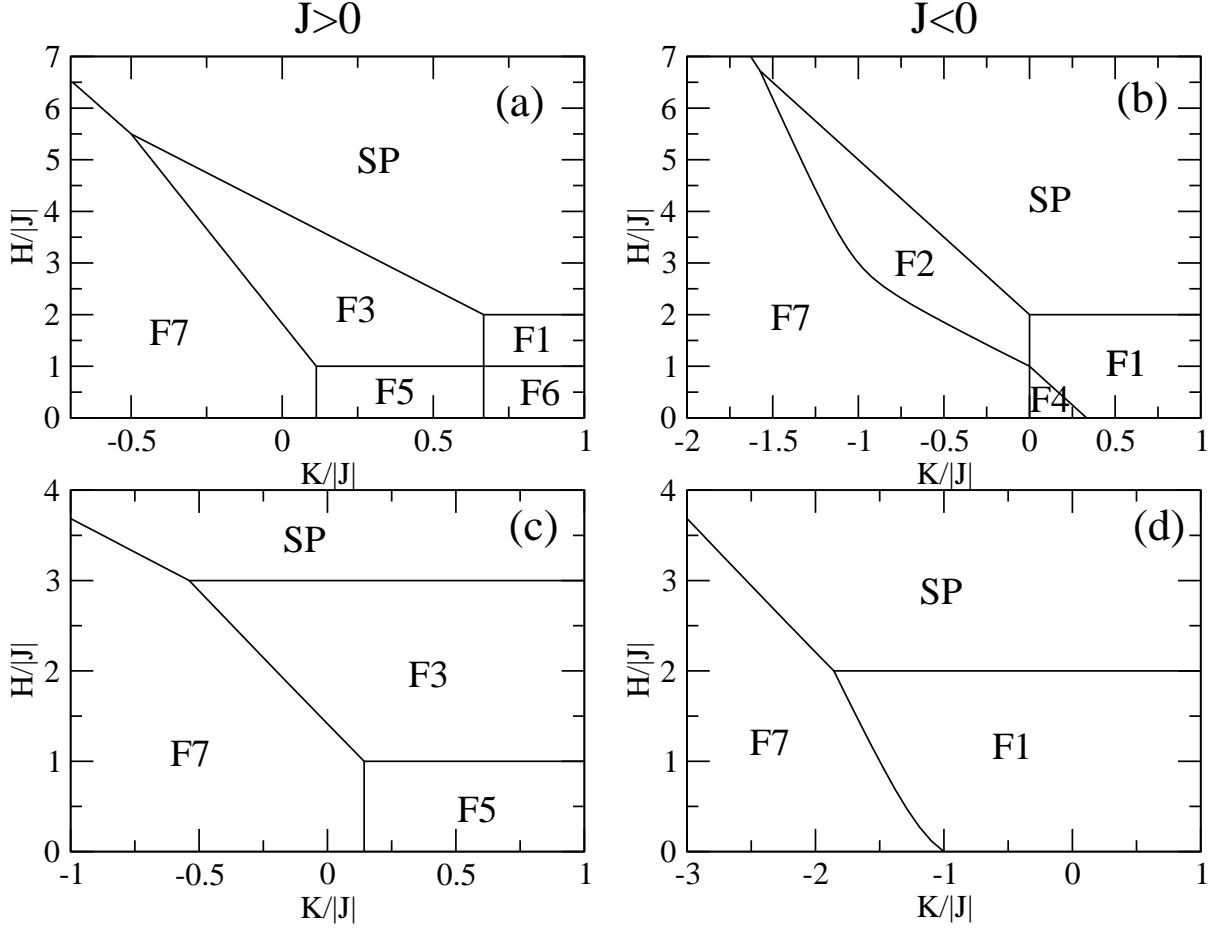}
\caption{\label{Fig-1}Ground states phase diagrams in ($\kappa$,$h$)-plane demonstrating the effect of biquadratic term.
Here $\kappa=K/|J|$ and $h=H/|J|$ The values of all other parameters, $j_0=J_0/|J|$, $\delta=D/|J|$ are
fixed as follows: $j_0=1/2, \delta=1/2$. Left panels ((a) and (c)) corresponds to the antiferromagnetic Heisenberg
interaction $J>0$, the right panels ((a) and (c)) - to ferromagnetic one $J<0$. In upper panels ((a) and (b)) $\Delta=2$
has been taken, while in lower panels ((c) and (d)) one can see phase diagrams for isotropic case $\Delta=1$.}
\end{figure}

Hereafter, to discuss the phase diagrams we will consider the
external magnetic field as $h_{1}$=$h_{2}$=$H$. It is also
convenient to present all parameters in the units of $|J|$. Thus, we
define $\kappa=K/|J|$, $j_0=J_0/|J|$, $\delta=D/|J|$ and $h=H/|J|$.
In Figure \ref{Fig-1} one can see four ground states phase diagrams
plotted in ($\kappa$,$h$)-plane demonstrating vast variety of ground
states for fixed values of the $\delta$ and $j_0$. These plots
summarize the effect of biquadratic term. The equations of phase
boundaries for $J>0, \Delta=2, j_0=0.5$ and $\delta=0.5$(Figure
\ref{Fig-1}(a)) are
\begin{align}
&\mbox{between} \quad  \mbox{F7}\quad\mbox{and} \quad \mbox{SP}, \quad &h=& \frac{1}{4}\left(6-15\kappa+\sqrt{\kappa^2+32 (\kappa-1)^2} \right), \\
&\mbox{between} \quad \mbox{F7} \quad\mbox{and} \quad \mbox{F3}, \quad &h=& \frac{1}{2}\left(-2-9\kappa+\sqrt{\kappa^2+32 (\kappa-1)^2} \right), \nonumber \\
&\mbox{between} \quad \mbox{F3} \quad\mbox{and} \quad \mbox{SP}, \quad &h=& 4-3\kappa. \label{PB1}\nonumber
\end{align}
The rest phase boundaries in this case are either horizontal or
vertical lines in ($\kappa$, $h$)-plane. $F5$ and $F3$ as well as
$F6$ and $F1$ are separated by the $h=1$ line, while the phase
boundary between $F5$ and $F6$ as well as between $F3$ and $F1$ are
vertical line is situated at $\kappa=2/3$. Ground states $F7$ and
$F5$ are separated by the line $\kappa=1/12(-17 + \sqrt{337})\approx
0.11313$.

For the case of antiferromagnetic Heisenberg interaction between $S=1$ spin ($J<0$) presented in Figure \ref{Fig-1}(b) the equations of the phase boundaries are
\begin{align}
&\mbox{between} \quad  \mbox{F7}\quad\mbox{and} \quad \mbox{SP}, \quad &h=&\frac{1}{4}\left(-15\kappa+\sqrt{32 (\kappa+1)^2+(\kappa+2)^2} \right), \\
&\mbox{between} \quad \mbox{F7} \quad\mbox{and} \quad \mbox{F2}, \quad &h=&\frac{1}{2}\left(-4-9\kappa+\sqrt{32 (\kappa+1)^2+(\kappa+2)^2} \right), \nonumber \\
&\mbox{between} \quad \mbox{F2}\quad \mbox{and} \quad \mbox{SP}, \quad &h=&2-3\kappa, \nonumber \\
&\mbox{between} \quad \mbox{F4}\quad  \mbox{and} \quad \mbox{F1}, \quad &h=&1-3\kappa. \label{PB2}\nonumber
\end{align}
The boundary between $F1$ and saturated ground state $SP$ is the straight line $h=2$. Horizontal line $\kappa=0$ separates $F4$ and $F7$ as well as $F1$ and $F2$. In order to demonstrate the significant role of exchange anisotropy $\Delta$ we plotted also the ground states phase diagrams for the isotropic case $\Delta=1$ ( Figure \ref{Fig-1}(c) and (d)) for antiferromagnetic and ferromagnetic $J$ respectively. One can see simplification of the ground state phase diagram via the disappearance of 2 ground states presented in the case anisotropic case $\Delta=2$( Figure \ref{Fig-1}(a) and (b)). So, for $J>0$ one can find, beside $SP$, only $F3$, $F5$, $F7$ and for $J<0$ only $F1$ and $F7$ ground states respectively. The equation of phase  boundaries for antiferromagnetic $J$, isotropic $\Delta=1$ and $j_0=0.5$, $\delta=0.5$ are:
\begin{align}
&\mbox{between} \quad  \mbox{F7}\quad\mbox{and} \quad \mbox{SP}, \quad &h=&\frac{1}{4}\left(6-3\kappa+\sqrt{8(\kappa-1)^2+\kappa^2} \right), \\
&\mbox{between} \quad \mbox{F7} \quad\mbox{and} \quad \mbox{F3}, \quad &h=&\frac{1}{2}\left(-\kappa+\sqrt{8(\kappa-1)^2+\kappa^2} \right), \nonumber
\end{align}
Ground states $F3$ and $SP$ are separated by the horizontal line $h=3$, another two straight lines appears between $F3$ and $F5$ and $F7$ and $F5$ at $h=1$ and $\kappa=1/7$ respectively. In the case of ferromagnetic $J$ and isotropic exchange interaction presented in Figure \ref{Fig-1}(d) one can see only three possible ground states $F1$, $F7$ and $SP$. Here the horizontal line $h=2$ separates $F1$ and $SP$ while other two phase boundaries are given by
 \begin{align}
&\mbox{between} \quad  \mbox{F7}\quad\mbox{and} \quad \mbox{SP}, \quad &h=&\frac{1}{4}\left(6-3\kappa+\sqrt{8(\kappa+1)^2+\kappa^2} \right), \\
&\mbox{between} \quad \mbox{F7} \quad\mbox{and} \quad \mbox{F1}, \quad &h=&\frac{1}{2}\left(2-3\kappa+\sqrt{8(\kappa+1)^2+\kappa^2} \right), \nonumber
\end{align}

\begin{figure}
\includegraphics[scale=0.5,angle=-90]{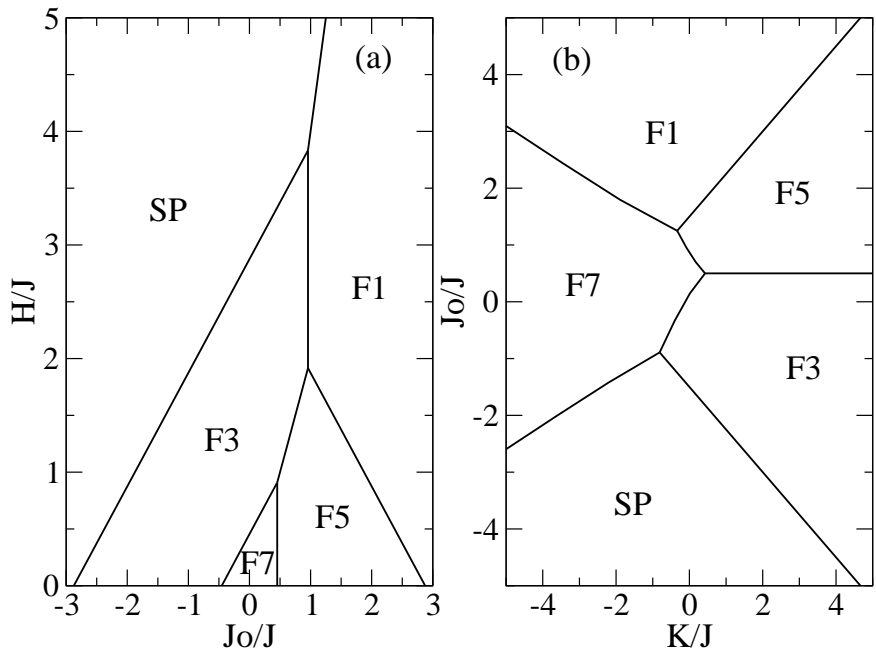}
\caption{\label{Fig-2}Ground states phase diagrams demonstrating the
influence of $J_0$. Here $\Delta=1/2$, $\kappa=1/2$ and $\delta=1$.
The panel (a) shows the ground states phase diagram in
($j_0$,$h$)-plane for fixed value of biquadratic interaction
$\kappa=1/2$; the panel (b) shows the ground states phase diagram in
($\kappa$, $j_0$)-plane for fixed value of magnetic field $h=1$.}
\end{figure}

\begin{figure}
\includegraphics[scale=0.5,angle=-90]{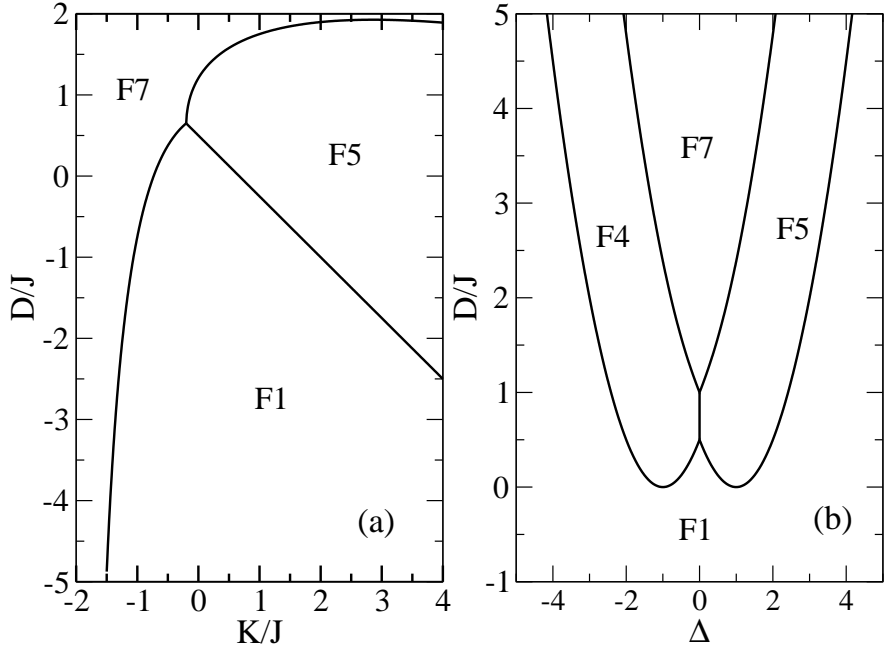}\caption{\label{Fig-3}Ground
states phase diagrams demonstrating the effect of single-ion anisotropy.
Here $h=1$, $j_0=1$. Panel (a) exhibit the ground states phase diagram in ($\kappa$, $\delta$)-plane
for fixed value of exchange anisotropy $\Delta=1/2$; panel (b) exhibits the ground states phase diagram
in ($\Delta$,$\delta$)-plane for fixed value of biquadratic interaction $\kappa=1/2$.}
\end{figure}

\begin{figure}
\includegraphics[angle=-90,scale=0.5]{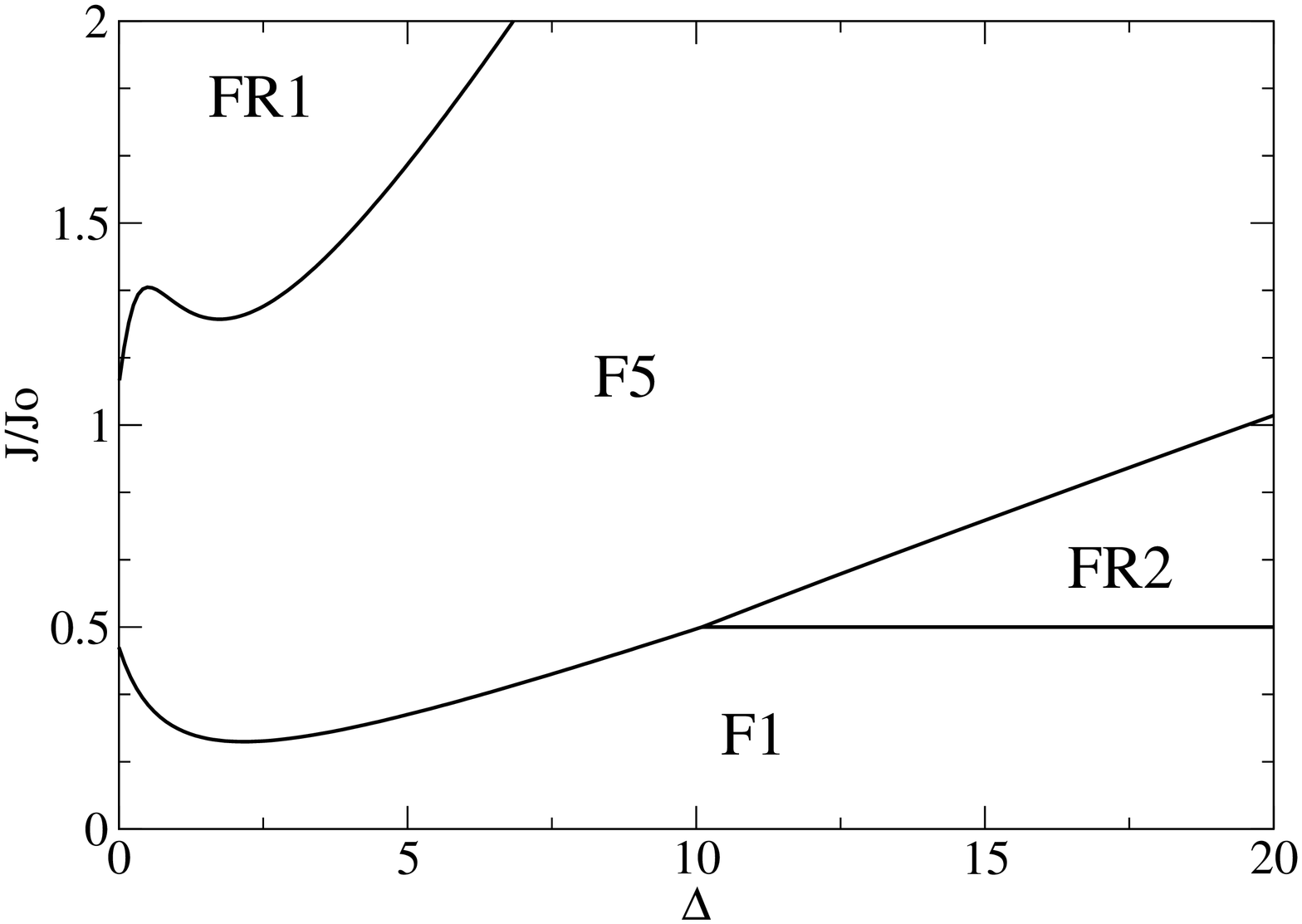}\caption{\label{fig:PDgm-Dlt-J}Zero field ground states
phase diagram drawn in the ($\Delta$, $J/J_{0}$)-plane for fixed $D=0$ and small values of $K/J_{0}=0.1$.
The frustrated ground states are exhibited.}
\end{figure}

In order ro summarize the effects of the Ising coupling $J_0$ we plotted another two ground state phase diagrams presented in Figure \ref{Fig-2}. Here, the left panel shows ground states boundaries in ($j_0$, $h$)-plane for fixed values of $\delta, \kappa$ and $\Delta$, while right panel demonstrate the phase boundaries for fixed  $\delta, h$ and $\Delta$ in ($\kappa$,$j_0$)-plane. For the sake of briefness we just list the equation of phase boundaries. For left panel:
\begin{align}
&\mbox{between} \quad  \mbox{F3}\quad\mbox{and} \quad \mbox{SP}, \quad &h=&\frac{23}{8}+j_0, \\
&\mbox{between} \quad \mbox{F3} \quad\mbox{and} \quad \mbox{F7}, \quad &h=&\frac{1}{2}\left(\frac{\sqrt{11}}{2}-\frac{3}{4} \right)+j_0, \nonumber\\
&\mbox{between} \quad  \mbox{F3}\quad\mbox{and} \quad \mbox{F5}, \quad &h=& 2 j_0, \nonumber \\
&\mbox{between} \quad \mbox{F5} \quad\mbox{and} \quad \mbox{F1}, \quad &h=&\frac{23}{8}-j_0, \nonumber\\
&\mbox{between} \quad  \mbox{F1}\quad\mbox{and} \quad \mbox{SP}, \quad &h=&4j_0, \nonumber \\
&\mbox{between} \quad \mbox{F7} \quad\mbox{and} \quad \mbox{F5}, \quad &j_0=&\frac{1}{2}\left(\frac{\sqrt{11}}{2}-\frac{3}{4} \right), \nonumber\\
&\mbox{between} \quad  \mbox{F3}\quad\mbox{and} \quad \mbox{F1}, \quad &j_0=&\frac{23}{24}, \nonumber
\end{align}
For right panel:
\begin{align}
&\mbox{between} \quad  \mbox{F7}\quad\mbox{and} \quad \mbox{SP}, \quad &j_0=&-\frac{1}{4}\left(1+\sqrt{2 (\kappa-1)^2+(\kappa+1)^2} \right), \\
&\mbox{between} \quad \mbox{F7} \quad\mbox{and} \quad \mbox{F3}, \quad &j_0=&\frac{1}{2}\left(2-\frac{3}{2}\kappa-\sqrt{2 (\kappa-1)^2+(\kappa+1)^2} \right), \nonumber\\
&\mbox{between} \quad  \mbox{F7}\quad\mbox{and} \quad \mbox{F5}, \quad &j_0=&\frac{1}{2}\left(-\frac{3}{2}\kappa+\sqrt{2 (\kappa-1)^2+(\kappa+1)^2} \right), \nonumber \\
&\mbox{between} \quad \mbox{F7} \quad\mbox{and} \quad \mbox{F1}, \quad &j_0=&\frac{1}{4}\left(3+\sqrt{2 (\kappa-1)^2+(\kappa+1)^2} \right), \nonumber\\
&\mbox{between} \quad  \mbox{SP}\quad\mbox{and} \quad \mbox{F3}, \quad &j_0=&-\frac{3}{2}\left(1+\frac{1}{2}\kappa \right), \nonumber \\
&\mbox{between} \quad \mbox{F3} \quad\mbox{and} \quad \mbox{F5}, \quad &j_0=&\frac{1}{2}, \nonumber\\
&\mbox{between} \quad  \mbox{F5}\quad\mbox{and} \quad \mbox{F1}, \quad &j_0=&\frac{3}{2}\left(1+\frac{1}{2}\kappa \right). \nonumber
\end{align}
Another two ground state phase diagrams demonstrating the influence
of single-ion anisotropy are presented in Figure \ref{Fig-3}.
Left(right) panel exhibits phase boundaries for for $h=1$, $j_0=1$
and $\Delta=0.5$ ($\kappa=0.5$) respectively. The phase boundaries
for the left and right panels respectively are
\begin{align}
&\mbox{between} \quad  \mbox{F1}\quad\mbox{and} \quad \mbox{F7}, \quad &\delta=&-\frac{3(2+2\kappa-\kappa^2)}{4 (2+\kappa)}, \\
&\mbox{between} \quad \mbox{F1} \quad\mbox{and} \quad \mbox{F5}, \quad &\delta=&\frac{1}{4}(2-\kappa), \nonumber\\
&\mbox{between} \quad  \mbox{F5}\quad\mbox{and} \quad \mbox{F7}, \quad &\delta=&\frac{1}{2}\left(1-\kappa+\sqrt{2+10\kappa+1/4\kappa^2} \right), \nonumber
\end{align}
and
\begin{align}
&\mbox{between} \quad  \mbox{F1}\quad\mbox{and} \quad \mbox{F4}, \quad &\delta=&\frac{1}{2}(\Delta+1)^2, \\
&\mbox{between} \quad \mbox{F1} \quad\mbox{and} \quad \mbox{F5}, \quad &\delta=&\frac{1}{2}(\Delta-1)^2, \nonumber\\
&\mbox{between} \quad  \mbox{F7}\quad\mbox{and} \quad \mbox{F4}, \quad &\delta=&\frac{1}{2}\left(1/2-\sqrt{9/4+\Delta^4-4 \Delta^3+5\Delta^2-6\Delta} \right), \nonumber\\
&\mbox{between} \quad  \mbox{F7}\quad\mbox{and} \quad \mbox{F5}, \quad &\delta=&\frac{1}{2}\left(1/2+\sqrt{9/4+\Delta^4+4 \Delta^3+5\Delta^2+6\Delta} \right), \nonumber\\
\end{align}
And finally, the zero field ground states phase diagram are
presented in Figure \ref{fig:PDgm-Dlt-J}. The phase diagram exhibits
the appearance of two frustrated ground states at zero magnetic
field. We choose $D=0$ and rather small value of biquadratic
interactions $K=0.1$. For convenience we normalize parameters by
$J_0>0$. The equations for the phase boundaries are
\begin{align}
&\mbox{between} \quad  \mbox{F1}\quad\mbox{and} \quad \mbox{F5}, \quad &\tilde{J}=&\frac{1-\tilde{K}(1-\Delta^2)}{1+\Delta}, \\
&\mbox{between} \quad \mbox{F1} \quad\mbox{and} \quad \mbox{FR2}, \quad &\tilde{J}=& 1, \nonumber\\
&\mbox{between} \quad  \mbox{FR2}\quad\mbox{and} \quad \mbox{F5}, \quad &\tilde{J}=&\frac{1+\tilde{K}(1-\Delta^2)}{1-\Delta}, \nonumber\\
&\mbox{between} \quad  \mbox{F5}\quad\mbox{and} \quad \mbox{FR1}, \quad &\tilde{J}=&\frac{2\Delta-1+\Delta (1+\Delta)(1+2\Delta)\tilde{K}+(1+\Delta(1+\Delta)\tilde{K})\sqrt{1+8 \Delta^2}}{2\Delta (1+\Delta)}, \nonumber\\
\end{align}
here $\tilde{J}=J/J_0$ and $\tilde{K}=K/J_0$.

\section{Exact Solution and Thermodynamics}
The present model could be solved exactly using the known decoration
transformation early presented by M. E. Fisher\cite{Fisher,syozi},
and recently generalized for arbitrary spin interactions\cite{physA388},
where one maps the partition function of the system to the partition
function of 1d Ising model, writing down the relations for the entries
of transfer matrix and thus obtaining the relations between model
parameters and that of Ising chain. But here we implement a direct
transfer matrix calculations without any account to the solution of
Ising chain. Therefore let us consider the following partition function
of the system, which can be represented as

\begin{equation}
\mathcal{Z}=\sum_{\sigma}\mbox{Sp}_{\mathbf{S}}\exp(-\beta\mathcal{H})=
\sum_{\sigma}\prod_{i=1}^{N}\exp(\beta\frac{h_{2}}{2}(\sigma_{i}+\sigma_{i+1}))Z(\sigma_{i},\sigma_{i+1}),\label{pf}\end{equation}
 where $\beta$ as usual is inverse temperature.
Here
\begin{equation}
Z(\sigma_{i},\sigma_{i+1})=\mbox{Sp}_{i}\exp(-\beta\mathcal{H}_{i})=\sum_{n=1}^{9}\exp(-\beta\lambda_{n}(\sigma_{i},\sigma_{i+1})).\label{z}
\end{equation}

Then, using \eqref{eig} one can express the one-block
\textit{partial} partition function $Z(\sigma_{i},\sigma_{i+1})$ in
the following form:
\begin{eqnarray}
Z(\sigma_{i},\sigma_{i+1}) & = & \sum_{n=0}^{2}Z_{n}\cosh\left(\beta n(h_{1}-J_{0}(\sigma_{i}+\sigma_{i+1}))\right),\\
Z_{0} & = & e^{\beta(J-K-2D)}+2e^{\beta\frac{1}{2}(J-(1+4\Delta^{2})K-2D)}\cosh\left(\tfrac{\beta R}{2}\right),\nonumber \\
Z_{1} & = & 4e^{-\beta(\Delta K+D)}\cosh\left(\beta\Delta J\right),\nonumber \\
Z_{2} & = & 2e^{-\beta(J+K+2D)}.\nonumber \end{eqnarray}
 After that, the partition function \eqref{pf} take the form
similar to the partition function of a chain with classical two
state variables on each site: \begin{eqnarray}
\mathcal{Z}=\sum_{\sigma}\prod_{i=1}^{N}T(\sigma_{i},\sigma_{i+1})=\mbox{Sp}\mathbf{T}^{N}=\Lambda_{1}^{N}+\Lambda_{2}^{N},\end{eqnarray}
 where $\Lambda_{1,2}$ are two eigenvalues of the transfer-matrix
\begin{eqnarray} \mathbf{T} & = &
\left(\begin{array}{lc}
e^{\beta\frac{h_{2}}{2}}\mathcal{Z}_{+} & \mathcal{Z}_{0}\\
\mathcal{Z}_{0} & e^{-\beta\frac{h_{2}}{2}}\mathcal{Z}_{-}\end{array}\right),\label{mat}\end{eqnarray}
 where \begin{eqnarray}
 &  & \mathcal{Z}_{\pm}=Z(\pm1/2,\pm1/2),\nonumber \\
 &  & \mathcal{Z}_{0}=Z(1/2,-1/2)=Z(-1/2,1/2).\end{eqnarray}
 Then, calculating the eigenvalues and taking thermodynamic limit,
when only the largest eigenvalue survives, one arrives at the following
expression for the free energy per block: \begin{eqnarray}
f=-\frac{1}{\beta}\log\left(\frac{1}{2}\left(e^{\beta\frac{h_{2}}{2}}\mathcal{Z}_{+}+e^{-\beta\frac{h_{2}}{2}}\mathcal{Z}_{-}+\sqrt{(e^{\beta\frac{h_{2}}{2}}\mathcal{Z}_{+}-e^{-\beta\frac{h_{2}}{2}}\mathcal{Z}_{-})^{2}+4{\mathcal{Z}_{0}}^{2}}\right)\right).\label{f}\end{eqnarray}
 As soon as the free energy per block is calculated, one can obtain
analytic expressions for all thermodynamic function.


\subsection{Magnetization and quadrupole moment}

Magnetic quantities can be obtained using the free energy expression
obtained in \eqref{f}. Therefore the magnetization of the spin-$S$
can be expressed as

\begin{equation}
M_{S}=\frac{1}{2N\mathcal{Z}}\sum_{\sigma}\mbox{Sp}_{\mathbf{S}}\left(\sum_{i=1}^{N}(S_{i1}^{z}+S_{i2}^{z})e^{-\beta\mathcal{H}}\right)=-\frac{1}{2}\left(\frac{\partial f}{\partial h_{1}}\right)_{\beta,h_{2},D},\end{equation}
 while the magnetization of spin-$\sigma$ reads

\begin{equation}
M_{\sigma}=\frac{1}{N/2\mathcal{Z}}\sum_{\sigma}\mbox{Sp}_{\mathbf{S}}\left(\sum_{i=1}^{N}\sigma_{i}e^{-\beta\mathcal{H}}\right)=-2\left(\frac{\partial f}{\partial h_{2}}\right)_{\beta,h_{1},D}.\end{equation}
 Thus, the total magnetization is given by

\begin{equation}
M=\frac{1}{5}M_{\sigma}+\frac{4}{5}M_{S}.\end{equation}

\begin{figure}
\includegraphics[scale=0.3]{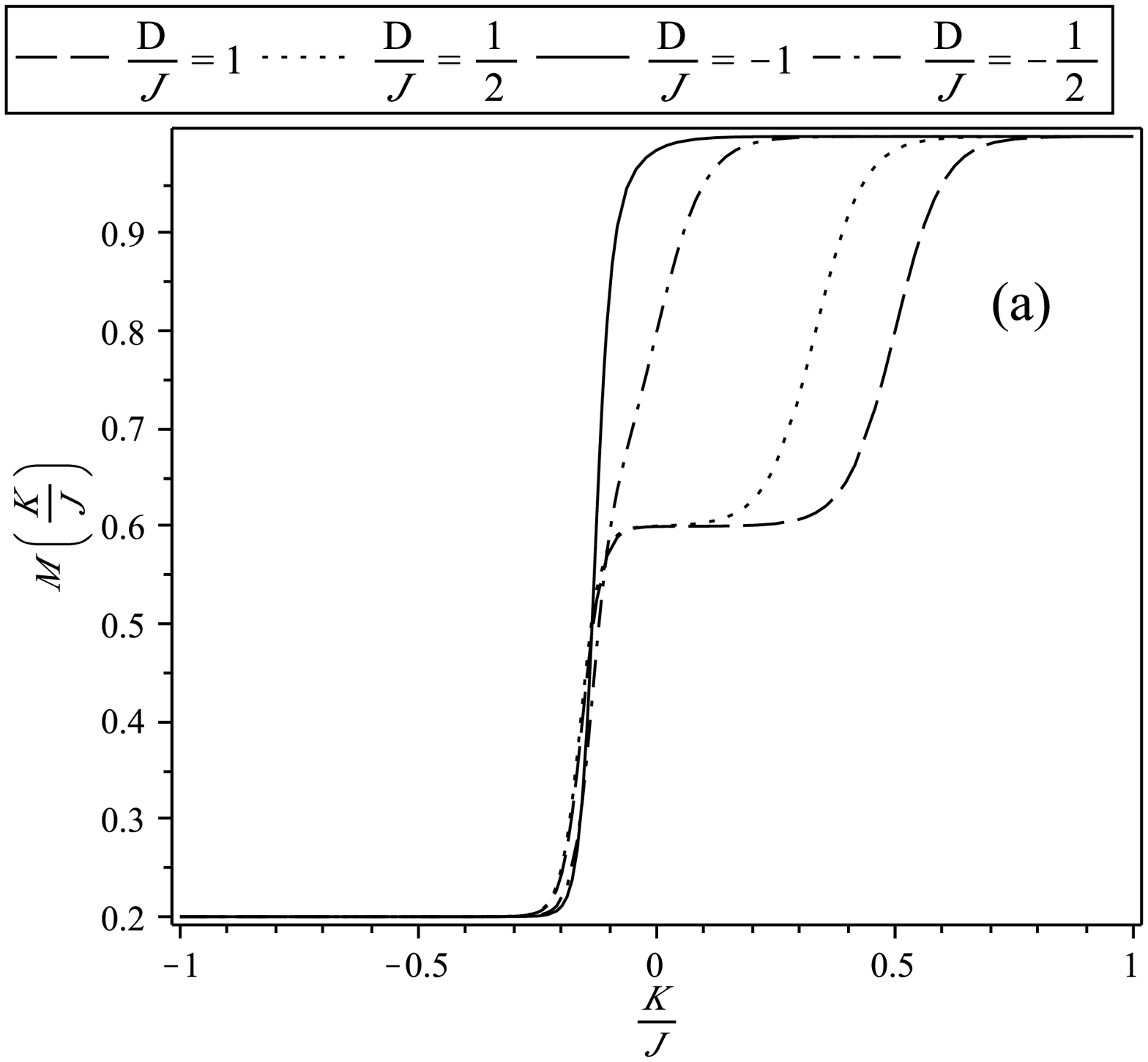}\includegraphics[scale=0.3]{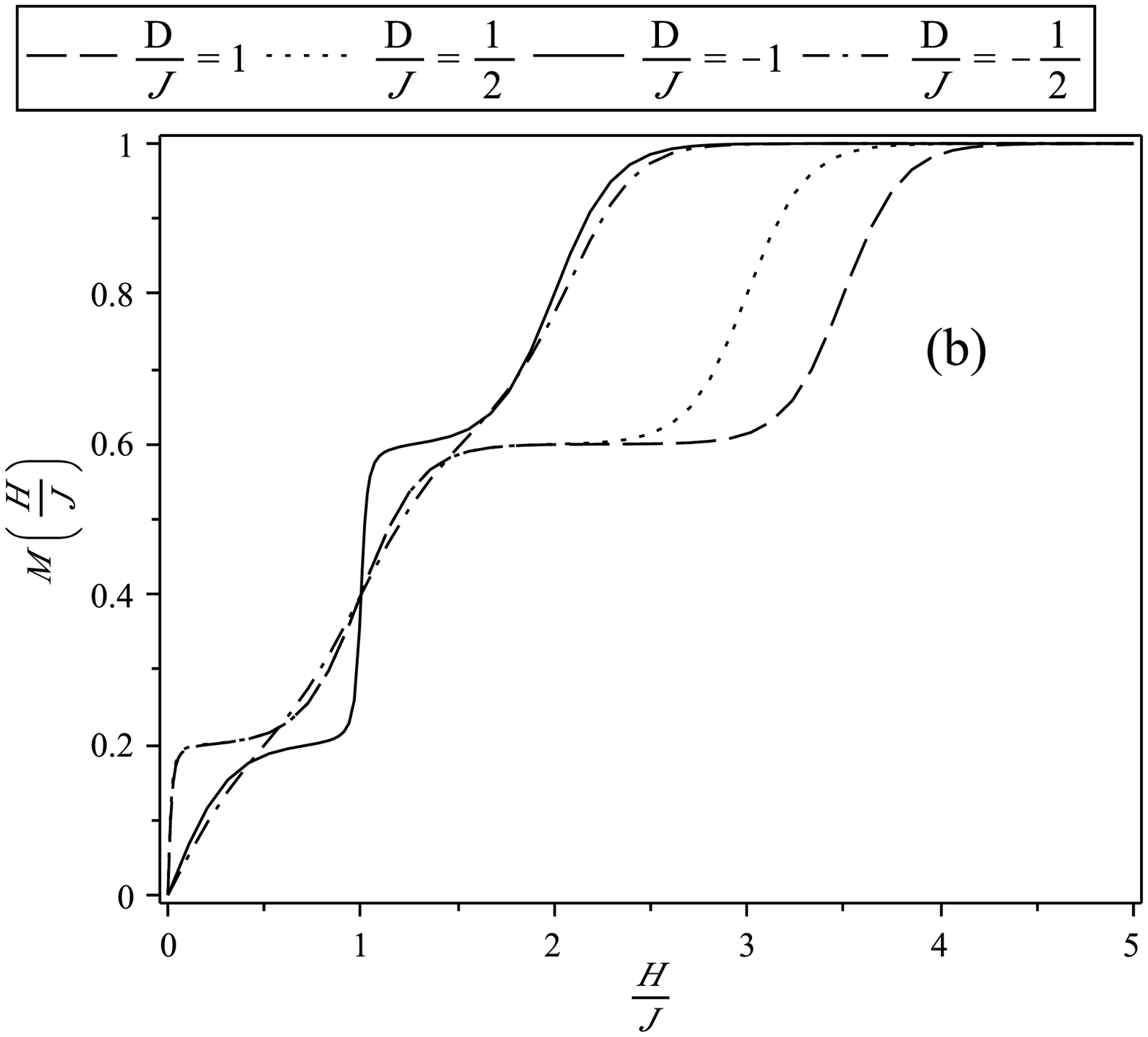}\includegraphics[scale=0.3]{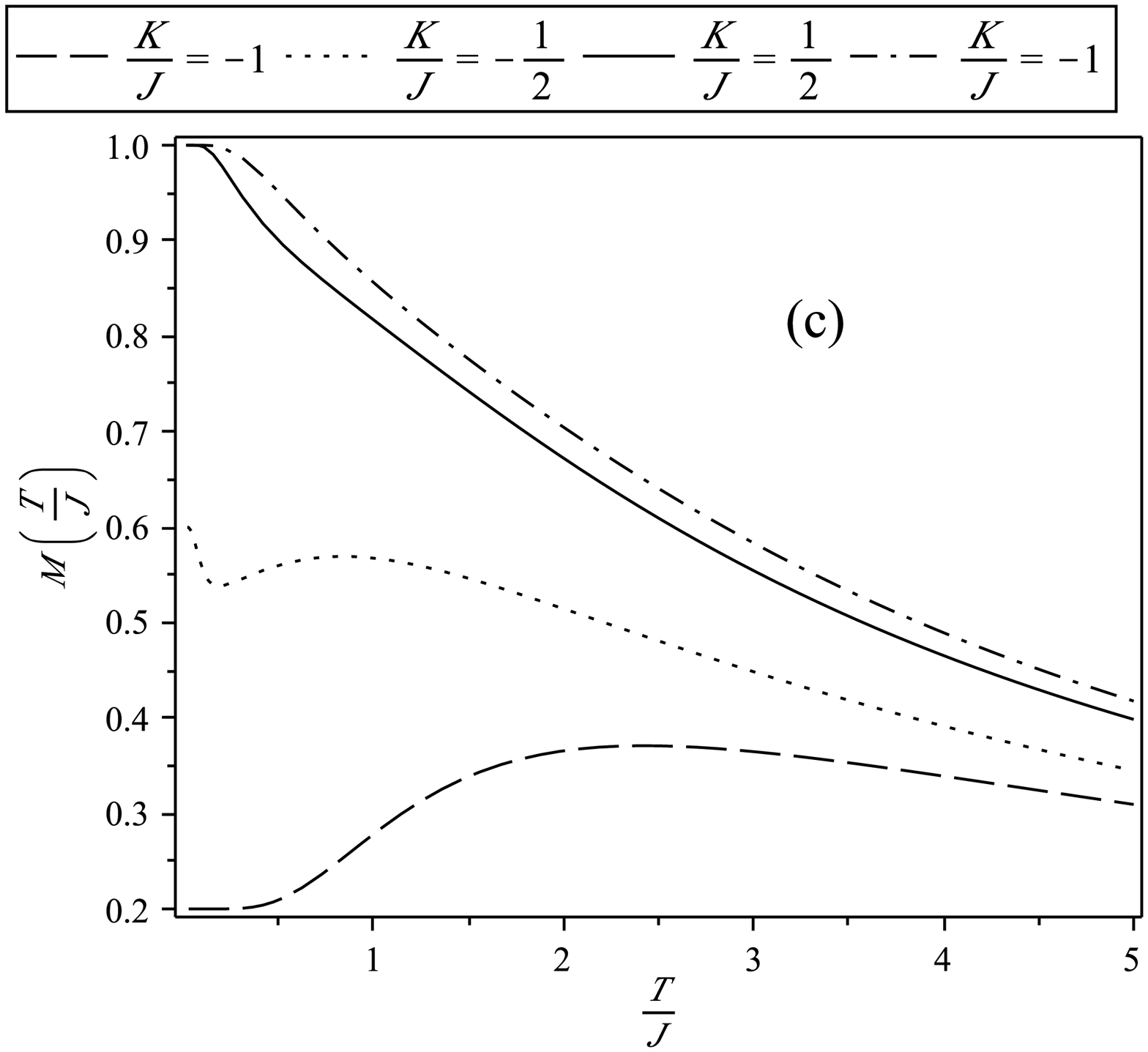}\caption{(a) Magnetization as a function of $K/J$, for $H/J=3.0$,
$T/J=0.15$, $\Delta=2$ and $J_0/J=0.5$. (b) Magnetization as a function
of $H/J$, for $K/J=0.5$, $T/J=0.15$, $\Delta=1$ and $J_0/J=0.5$.
(c) Magnetization versus temperature $T/J$, for $J_0/J=0.5$, $H/J=4$, $\Delta=1.4$ and $D/J=1$ .\label{fig:4}}

\end{figure}

Figure \ref{fig:4}(a) displays the plot of magnetization as a
function of biquadratic interaction $K$ in units of $J$, for fixed
values of $H/J=3$, $T/J=0.15$, $\Delta=2$ and $J_0/J=0.5$. Here one
can see the plateaus for different values of $J$, these plateaus
occur as expected at 1/5 and 3/5. The plots of magnetization
processes ($M$ versus $H$) for the system under consideration is
presented in figure \ref{fig:4}(b). Here the values of parameters
are fixed as $K/J=0.5$, $T/J=0.15$, $\Delta=1$  and the value of
$D/J$ varies from -1 to 1. As in can see from the phase diagram in
Figure (\ref{Fig-1})(c), two intermediate magnetization plateaus at
$M=1/5$ and $M=3/5$ exhibited by the system in the case correspond
to $F3$ and $F5$ ground sates for $T=0$.
 Thermal behavior of magnetization at the fixed external field
is presented in figure \ref{fig:4}(c) where the following values of
parameters are assumed:  $H/J=4$, $\Delta=1.4$, $D/J_0=1$ and
$J_0/J=0.5$, the competition between ferromagnetic state $SP$ and
ferrimagnetic state $F1$ with magnetization $M=3/5$ is occurred with
increase of the temperature. Close to zero temperature one obtains
three well defined values for the magnetization, which are in
accordance with plateaus displayed in figure \ref{fig:4}(a-b).


%

As the system under consideration contains sites with spin-1 one can
define another important physical quantity, quadrupole moment, which
can be obtained by the thermodynamic relations as well

\begin{equation}
Q=\frac{1}{2N\mathcal{Z}}\sum_{\sigma}\mbox{Sp}_{\mathbf{S}}\left(\sum_{i=1}^{N}((S_{i1}^{z})^{2}+(S_{i2}^{z})^{2})e^{-\beta\mathcal{H}}\right)=\frac{1}{2}\left(\frac{\partial f}{\partial D}\right)_{\beta,h_{1},h_{2}}.\end{equation}

\begin{figure}
\includegraphics[scale=0.3]{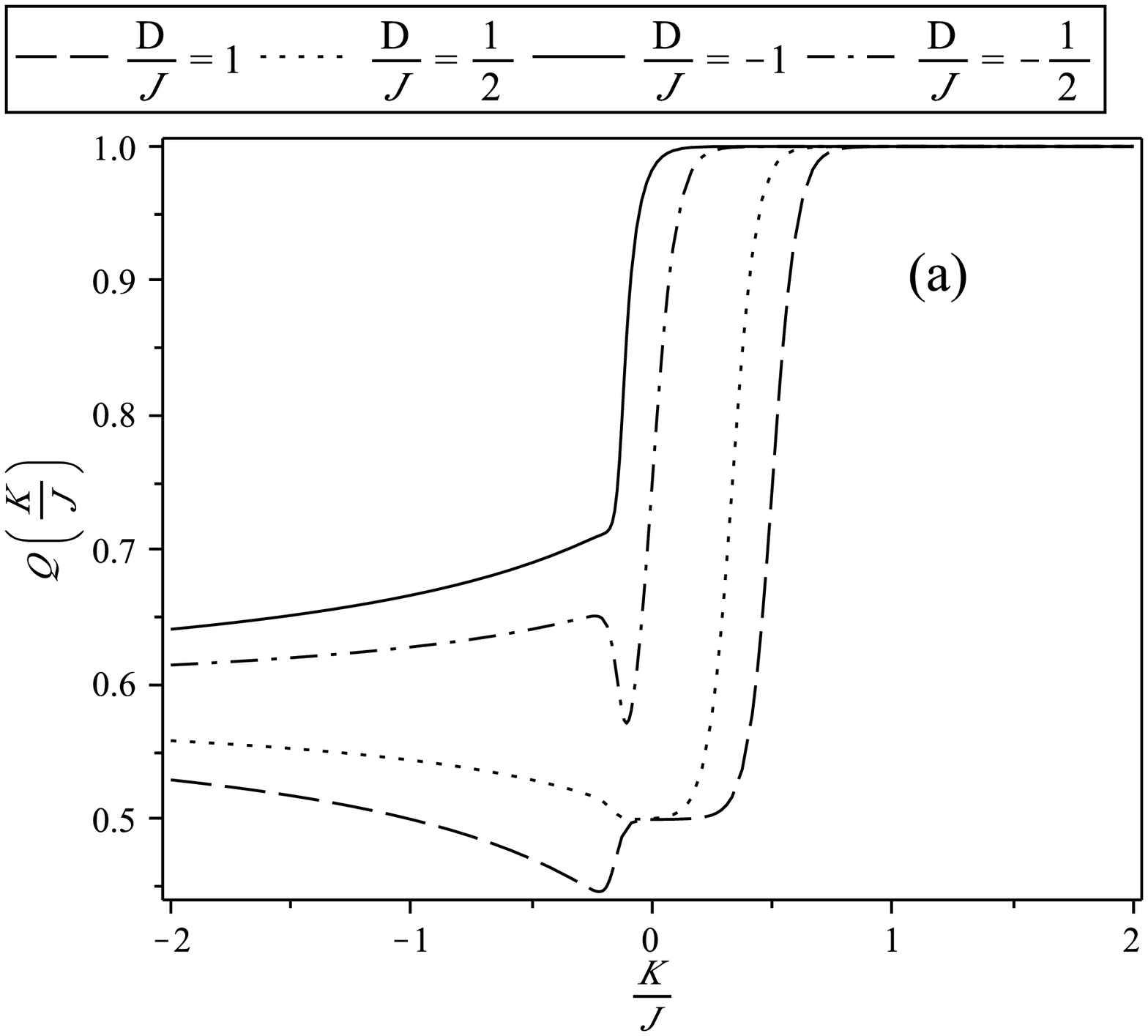}\includegraphics[scale=0.32]{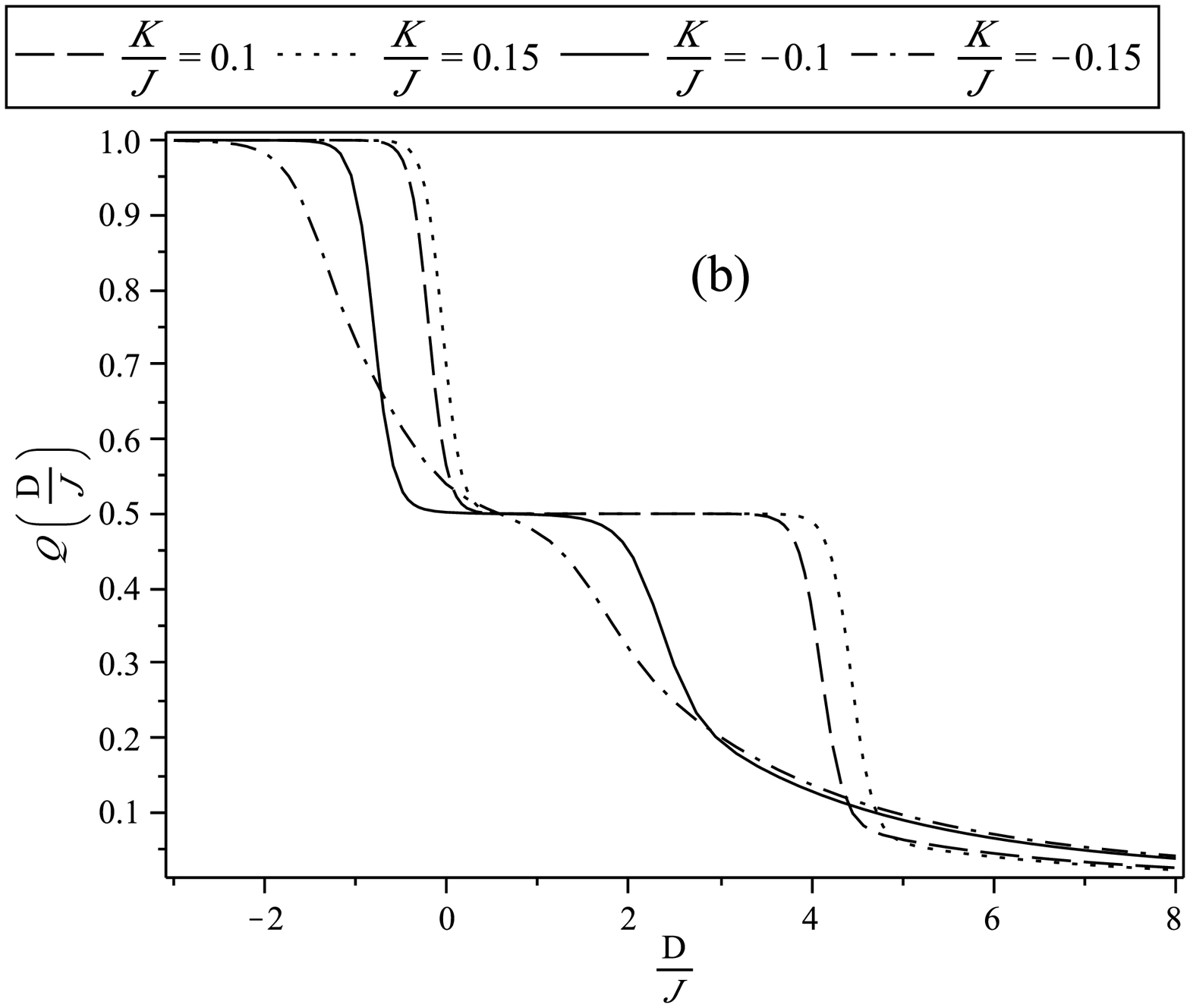}\includegraphics[scale=0.3]{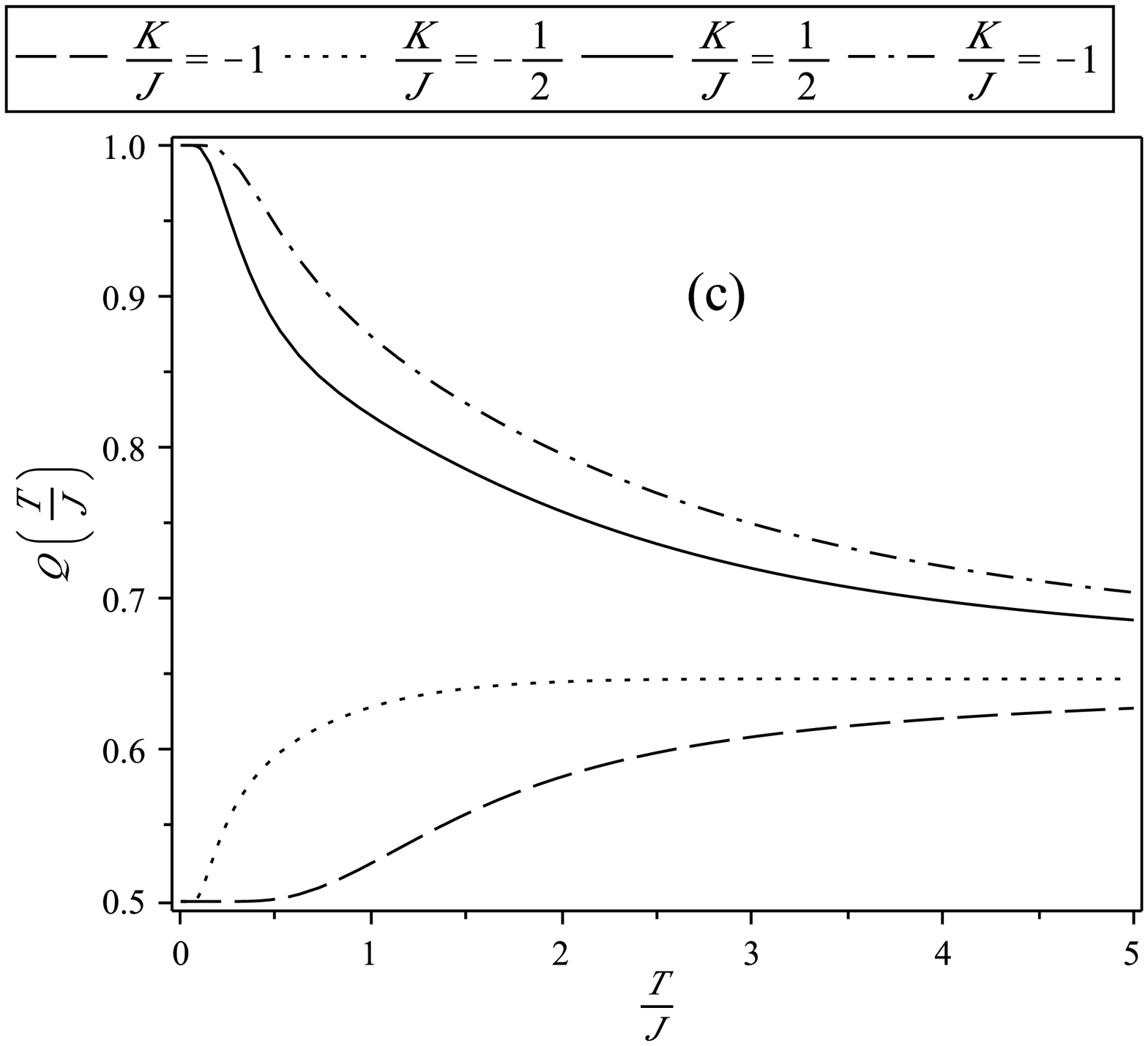}\caption{Quadrupole Moment: (a) As a function of $K/J$, for $H/J=3$, $J_0/J=0.5$, $\Delta=2$ and $T/J=0.15$. (b) As a function of $D$, for
$J/J_0=-0.5$, $T/J=0.1$, $H/J=2$ and $\Delta=2$. (c) As
a function of $T$, for $J_0/J=0.5$, $D/J=1.0$, $H/J=4$ and
$\Delta=1.4$.\label{fig:5}}

\end{figure}

In figure \ref{fig:5}(a) the plots of quadrupole moment as a
function of $K/J$ for fixed values of $H/J=3$, $J_0/J=0.5$,
$\Delta=2$ and $T/J=0.15$ are presented for several temperatures.
The non-trivial and non-monotone behavior of $Q$ under variation of
$K$ can be understood if one take into account appearance of $F6$
and $FR2$ ground states in which vertical quantum dimer is in
$|v_{0,-}\rangle$ eigenstate. Calculating expectation value for the
operator Q for this state one obtains: \begin{eqnarray} \langle
v_{0,-}|\frac{1}{2}((S_{1}^{z})^{2}+(S_{2}^{z})^{2})|v_{0,-}\rangle=\frac{1}{1+\frac{1}{8\Delta^{2}}\left(1+\frac{2D-R}{K-J}\right)^{2}},\label{eq:Q}\end{eqnarray}
 which actually defines low-temperature behavior of the quadrupole
moment. 
The quadrupole moment dependence of the uniaxial single-ion
anisotropy parameter $D$ is illustrated in figure \ref{fig:5}(b) for
$H/J=2$, $J_0/J=-0.5$, $T/J=0.1$ and $\Delta=2$. Similar to the
magnetization case, we obtain some plateaus, but for higher values
of $D$ we have a decreasing curve instead of plateaus(solid line).
On the other hand, as soon as the temperature increases these
plateaus obviously disappear. In the Figure \ref{fig:5}(c) we show
the quadrupole moment as a function of the temperature for fixed
values of $J_0/J=0.5$, $D/J=1.0$, $H/J=4$ and $\Delta=1.4$, similar
to the case of magnetization (fig.\ref{fig:4}(c)), the quadrupole
moment leads to fixed values at low temperature which are related to
the plateaus displayed in figures \ref{fig:5}(a-b), while at high
temperature average quadrupole moment leads to $2/3$, which
corresponds to equal probabilities for all three values of $S=1$
spin.

\subsection{Entropy and specific heat}

The entropy of the system can be obtained according to general thermodynamic
relation,

\begin{equation}
\mathcal{S}(T)=-\left(\frac{\partial f}{\partial T}\right)_{h_{1},h_{2},D}.\end{equation}

\begin{figure}
\includegraphics[scale=0.3]{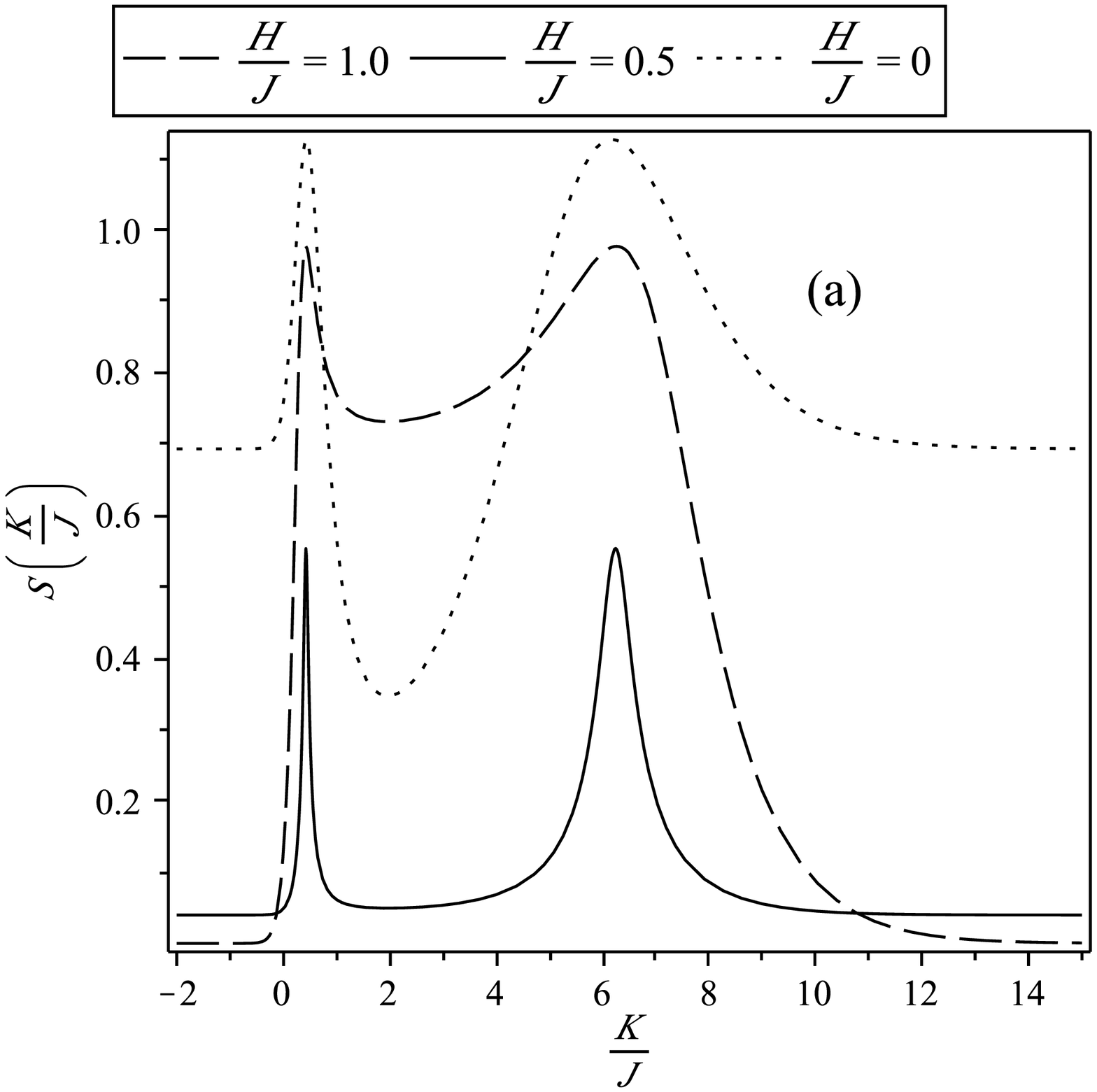}\includegraphics[scale=0.3]{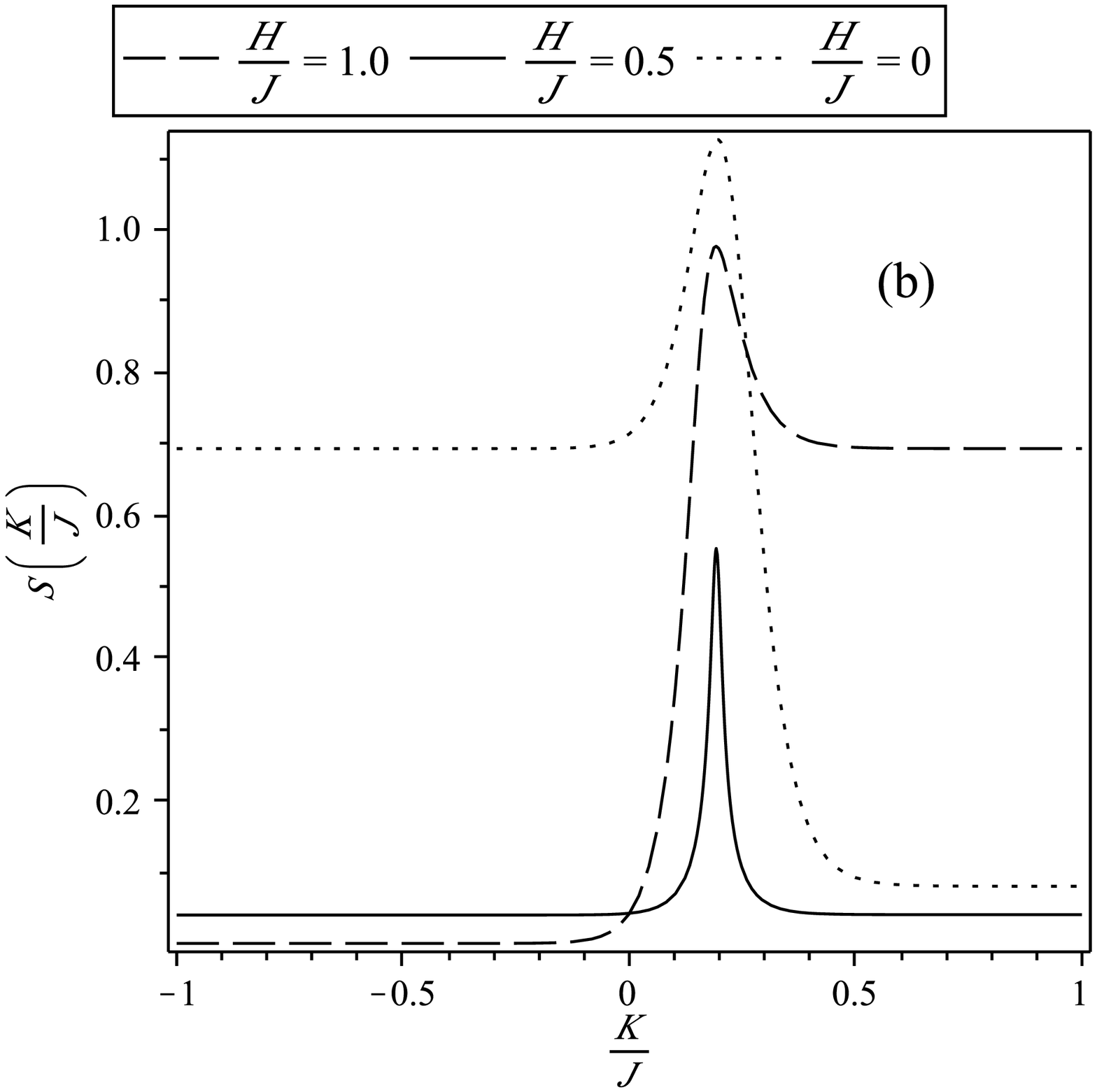}
\includegraphics[scale=0.3]{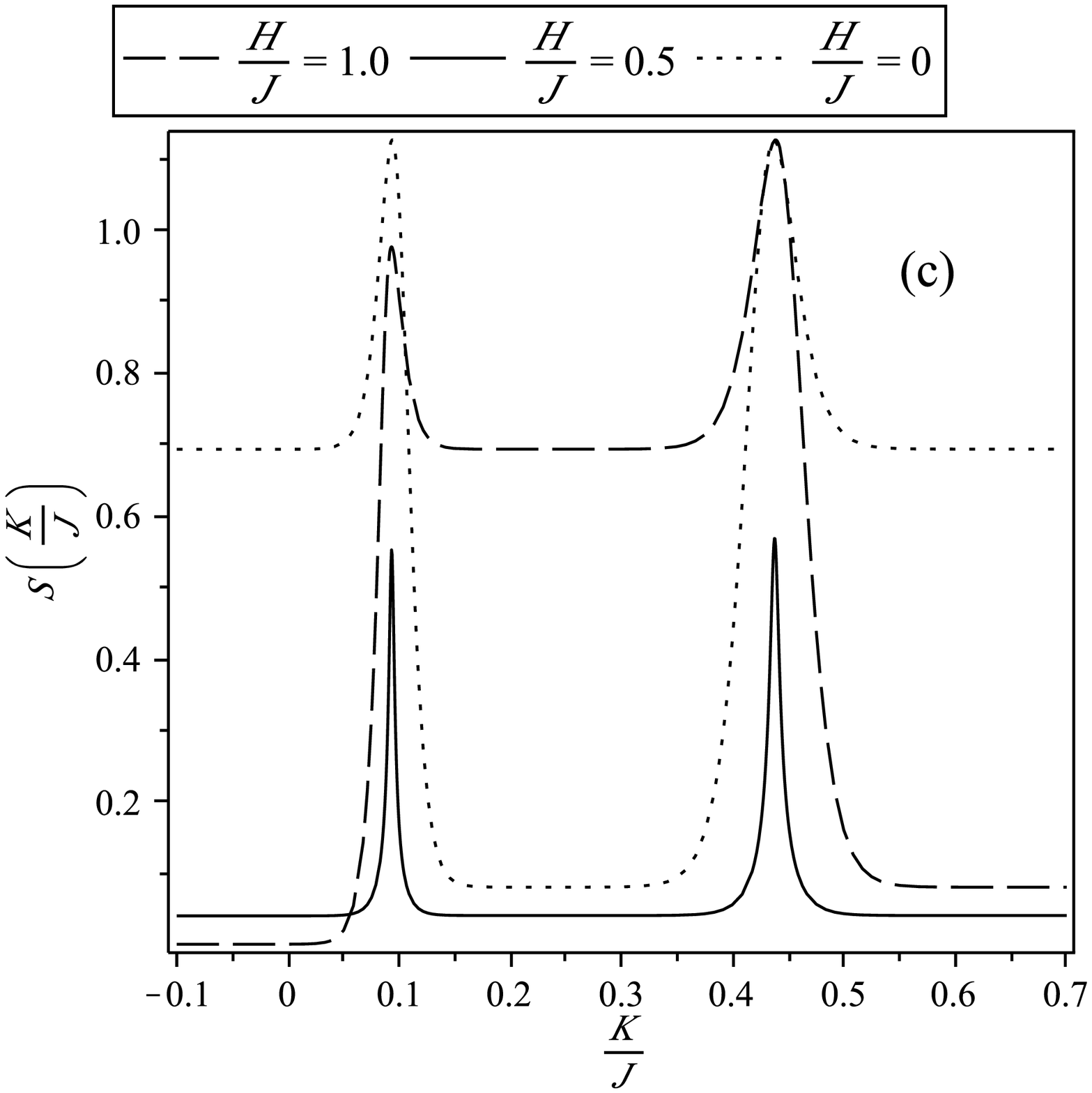}\caption{\label{fig:6}The entropy $\mathcal{S}(T)$ against $J$: assuming fixed
values for  $J_{0}/J=0.5$, $T/J=0.1$, $D/J=1$.  (a)  $\Delta=0.5$, (b) $\Delta=1$, (c) $\Delta=3$.}

\end{figure}

One can see the plot of entropy $\mathcal{S}(T)$ of the model as a
function of $K/J$ in figure \ref{fig:6}, here we fix the following
values of the parameters: $J_{0}/J=0.5$, $T/J=0.1$, $\Delta=3$ and
$D/J=1$, assuming the external magnetic field is fixed at $H/J=1.0$,
0.5 and 0.1. Three panels correspond to three values of $\Delta$,
(a) $\Delta=0.5$, (b) $\Delta=1$, (c) $\Delta=3$.
 A series of
peaks corresponding to quantum phase transition points appear on the
curve which is in accordance with the phase transition at zero
temperature illustrated in figure \ref{Fig-1}. According to previous
discussion, when external magnetic field vanishes there are a
residual entropy of the model corresponding to frustrated ground
states. This residual entropy is a constant value
$\mathcal{S}(0)=\ln(2)$=0.693 related to the ground state
degeneracy. For particular values of the magnetic field like $H/J=1$
it is possible to occur a twofold degeneracy of the ground state
energy.

Using the entropy $\mathcal{S}(T)$, one can obtain the specific heat

\begin{equation}
C=T\left(\frac{\partial\mathcal{S}}{\partial
T}\right)_H.\end{equation}

\begin{figure}
\includegraphics[scale=0.3]{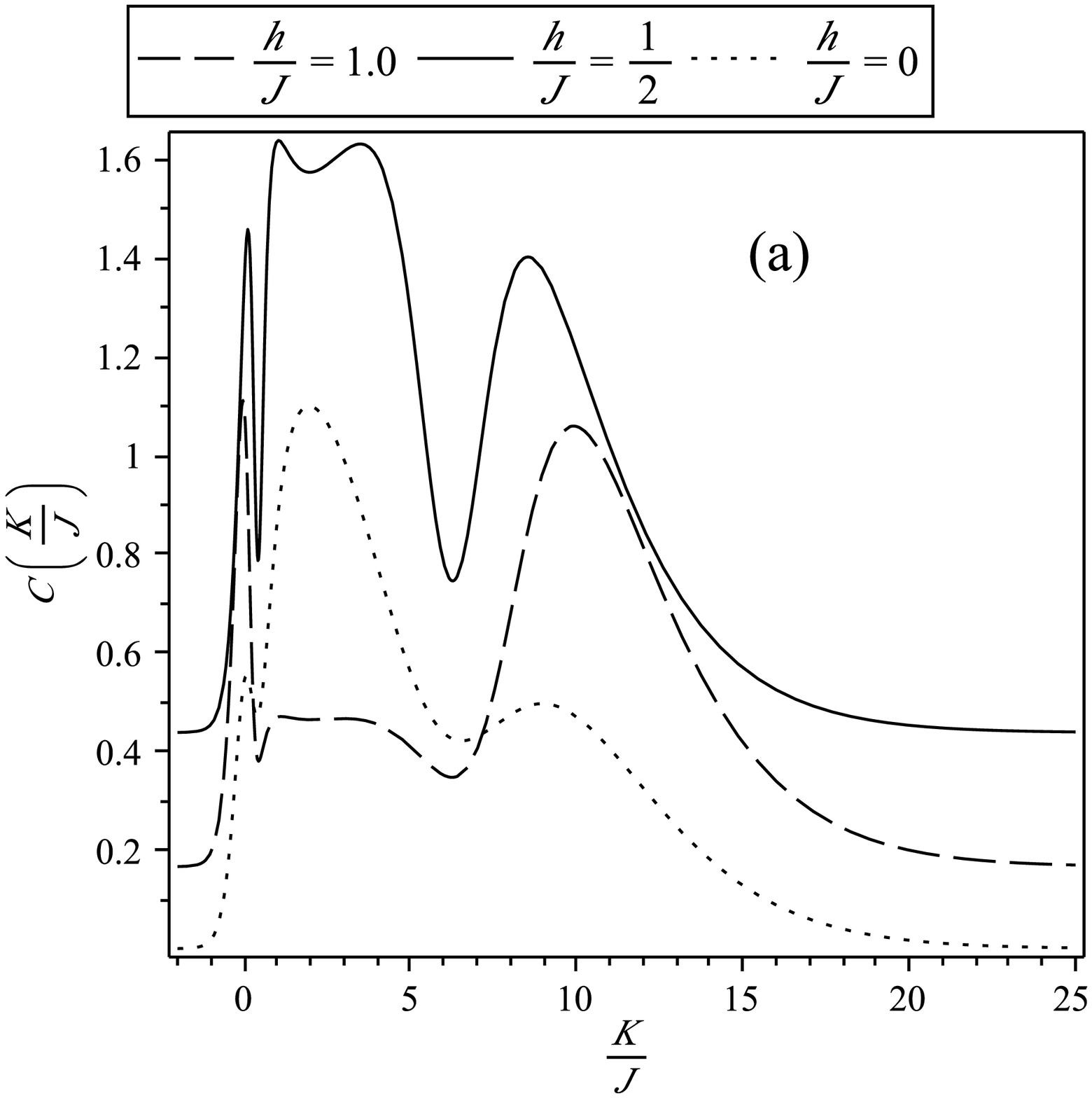}\includegraphics[scale=0.3]{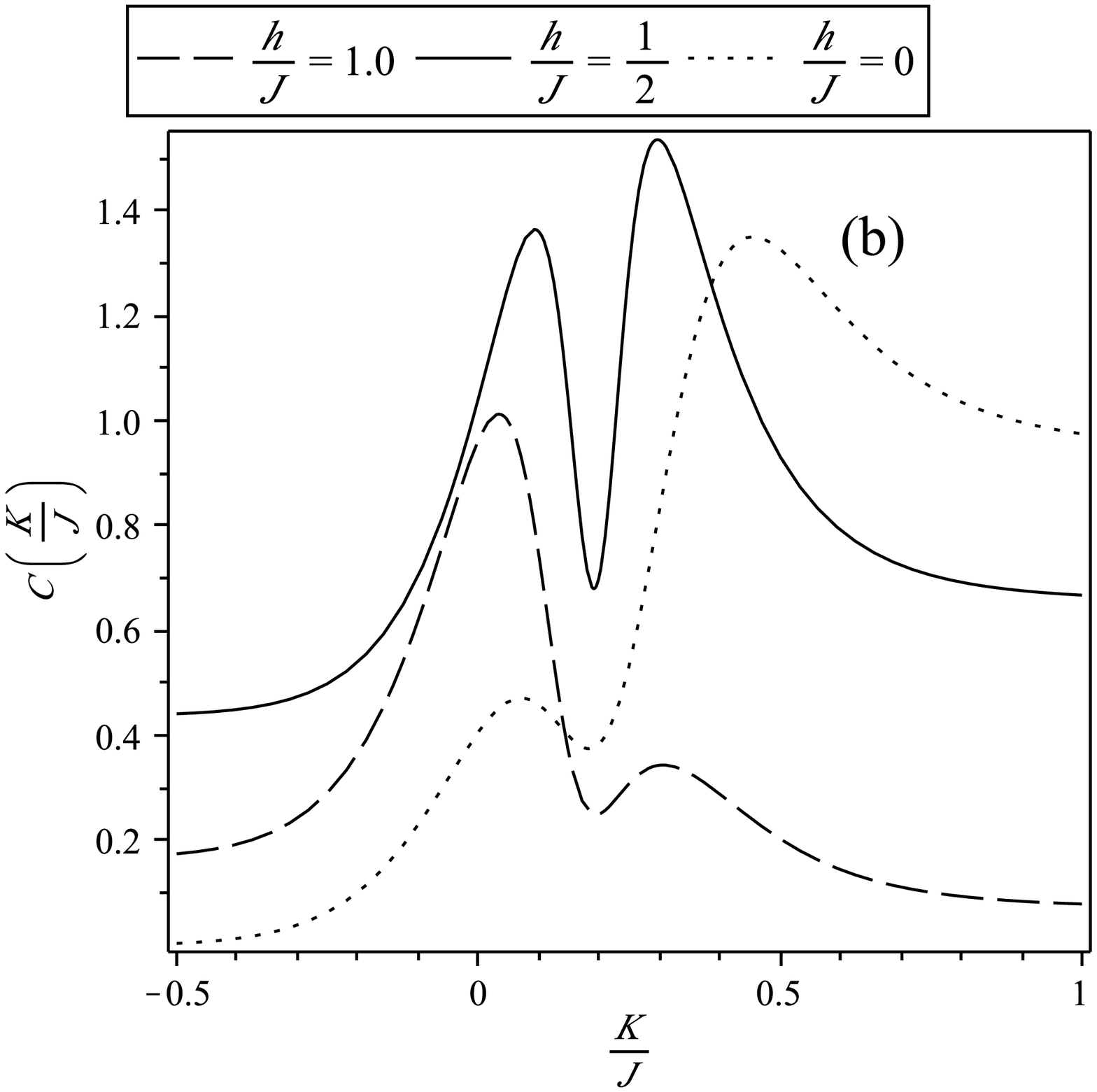}\includegraphics[scale=0.3]{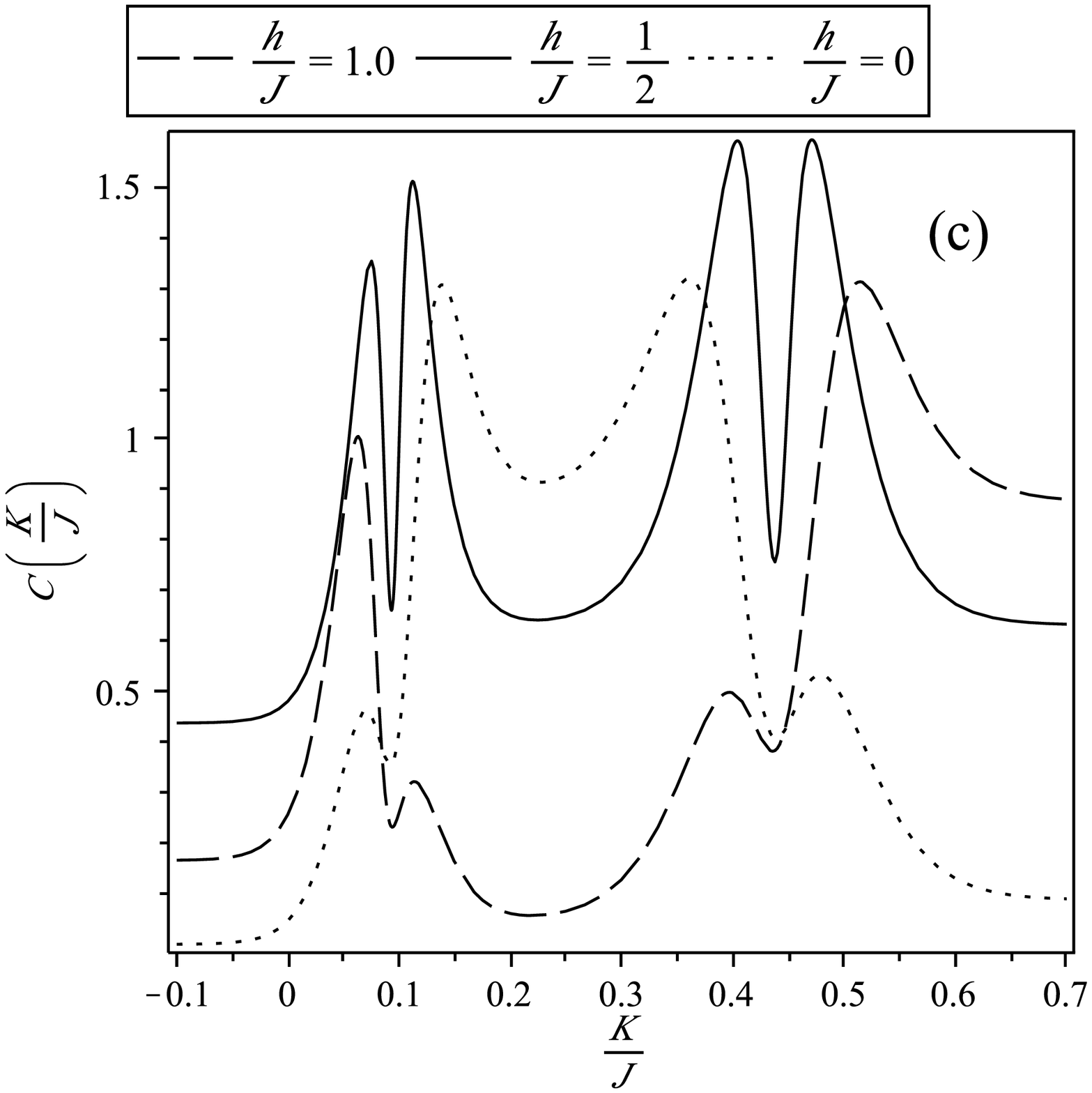}
\caption{\label{fig:7}The specific heat $C(T)$,
for the fixed values of $J_0/J=0.5$, $T/J=0.2$, $J_{0}/J=-0.5$ and $D/J=1$.  (a)  $\Delta=0.5$, (b) $\Delta=1$, (c) $\Delta=3$.}
\end{figure}

In order to study the specific heat properties for the model under
consideration, in fig.\ref{fig:7}  we display the specific heat as a
function of the parameter $K/J$ for the fixed values of $J_0/J=0.5$,
$T/J=0.2$ and $D/J=1$. For three values of $\Delta$, (a)
$\Delta=0.5$, (b) $\Delta=1$, (c) $\Delta=3$. In figure \ref{fig:7}
the corresponding plots exhibit four peaks which are related to the
phase transition at zero temperature. As one can see the phase
transition effects appears for small values of the biquadratic term
$K$.

\section{Conclusions}

We have considered an exactly solvable variant of diamond chain with
mixed $S=1$ and $S=1/2$ spins. The vertical $S=1$ dimers are taken
as quantum ones with Heisenberg bilinear and biquadratic interactions
and with single-ion anisotropy terms, while all interactions between
$S=1$ spins and $S=1/2$ spins residing on the intermediate sites
are taken in the Ising form. The model generalizes the model of diamond
chain with Ising and Heisenberg bonds considered in \cite{str05}.
Our results supplement the previously obtained ones for the case of
$S=1$ vertical XXZ-dimers with only bilinear Heisenberg interaction.
The detailed analysis of the $T=0$ ground state phase diagrams is
presented as well as the exact plots of various thermodynamic functions.
The effect of biquadratic term and single-ion anisotropy are summarized
in the corresponding phase diagrams.
The phase diagrams have shown to be rather rich, demonstrating large
variety of ground states: saturated one, three ferrimagnetic with
magnetization equal to 3/5 and another four ferrimagnetic ground states
with magnetization equal to 1/5. There are also two frustrated macroscopically
degenerated ground states which could exist at zero magnetic filed.
The thermodynamic properties of the model have been described exactly
by exact calculation of partition function within the direct classical
transfer-matrix formalism, the entries of transfer matrix, in their
turn, contain the information about quantum states of vertical $S=1$
XXZ dimer (eigenvalues of local hamiltonian for vertical link). The
plots of entropy and specific heat are also presented.

\section*{Acknowledgments}

V.O. express his gratitude to Institut f\"ur Theoretische Physik Universit\"at
G\"ottingen for warm hospitality during the final stage of this work.
This research stay was supported by DFG (Project No. HO 2325/7-1).
V.O and M. K. were partly supported by the joint grant of CRDF-NFSAT and
State Committee of Science of Republic of Armenia ECSP-09-94-SASP,
the work of V. O. was also partly supported by Volkswagen Foundation
(grant No. I/84 496) and ANSEF-1981-PS. O.R. and S.M.S thanks CNPq
and FAPEMIG for partial financial support.

\end{document}